\documentclass[11pt,a4paper]{article}
\usepackage[T1]{fontenc}
\usepackage{jheppub}
\counterwithout{equation}{section}

\makeatletter
\let\@fpheader\@empty
\makeatother
\usepackage{mathrsfs,bm,booktabs}
\usepackage{microtype}
\hypersetup{
 pdftitle={Entropy and charge transport limit Mpemba preparation in holography},
 pdfauthor={Yu Shi, Sheng Long}
}
\graphicspath{{figures/}}
\newcounter{figurepanel}[figure]

\newcommand{\manuscriptpanel}[4][.5]{%
 \begin{minipage}[b]{#1\linewidth}
  \raggedright
  \hspace*{#2\linewidth}\refstepcounter{figurepanel}\label{#4}%
  {\fontsize{9}{10}\selectfont(\alph{figurepanel})}\par
  \nointerlineskip
  \includegraphics[width=\linewidth]{#3}%
 \end{minipage}%
}
\newcommand{\ii}{\mathrm{i}}
\newcommand{\dd}{\mathrm{d}}
\newcommand{\Real}{\operatorname{Re}}
\newcommand{\Imag}{\operatorname{Im}}

\newcommand{\cN}{\mathcal N}
\title{Entropy and charge transport limit Mpemba preparation in holography}
\author[a]{Yu Shi}
\author[b,1]{and Sheng Long\note{Corresponding author.}}
\affiliation[a]{Department of Physics, Huaihua University, Huaihua, Hunan 418008, China}
\affiliation[b]{School of Fundamental Physics and Mathematical Sciences, Hangzhou Institute for Advanced Study, University of Chinese Academy of Sciences, Hangzhou 310024, China}
\emailAdd{shiyu@hhtc.edu.cn}
\emailAdd{shenglong@ucas.ac.cn}
\arxivnumber{2611.15654}

\abstract{
We study Mpemba relaxation and its preparation cost in a holographic
superconductor with full backreaction. At matched final energy and charge,
boundary drives of finite duration can reverse the ordering of the remaining
horizon entropy after source removal. In separate comparisons, some protocols
reach a prescribed entropy tolerance earlier even when driving and waiting
times are included. At criticality, a slow scalar field can persist even when
little horizon entropy remains available for preparation and subsequent
relaxation. Along the critical slow branch at fixed energy and charge,
horizon displacement cancels the quadratic area response, leaving an entropy
reserve of order $\varepsilon^4$ for an initial scalar amplitude
$\varepsilon$. The same coefficient governs the entropy advantage of the
nearby superconducting equilibrium and the entropy weight of the slow mode
on the normal side. This weight vanishes in proportion to the mode's decay
rate near the transition. Horizon area growth is quadratic in the covariant
scalar flux. When the horizon remains smooth and its area growth uniformly
controls the radial drift of the scalar field, the time needed to suppress the
leading horizon field and exterior charge grows at least as
$\varepsilon^{-2}$. Charge transfer fixes the signed area swept out by the
time integral of this same flux, raising the minimum flux cost and
strengthening the time bound.
}

\begin{document}
\maketitle
\flushbottom

\section{Introduction}
\label{sec:intro}

In the Mpemba effect, a state that starts farther from equilibrium can
approach it sooner than one that starts closer~\cite{Mpemba1969,Bechhoefer2021,Teza2026,LuRaz2017}.
This reversal is possible because the approach to equilibrium depends both
on the decay rate of each mode and on how strongly the initial state excites it.
Suppressing the slowest mode can accelerate relaxation, while eliminating its
asymptotic amplitude gives the strong Mpemba effect~\cite{Klich2019}. This
possibility has been explored in experiments with colloids and trapped
ions~\cite{Kumar2020,Zhang2025,AharonyShapira2024} and in studies of quantum
state preparation~\cite{Carollo2021,Kochsiek2022,Ivander2023,BeatoTeza2026,AresCalabreseMurciano2025Review,Lejeune2026}.
Related relaxation phenomena have also been studied using entanglement
asymmetry~\cite{Ares2023,Rylands2024,Joshi2024}. Whether suppressing a slow
mode saves time overall also depends on how the state is prepared.

Preparing a target state typically involves a drive of finite duration that
changes the state at release and can produce dissipation~\cite{GueryOdelin2023}.
In driven granular fluids, the velocity distributions that can be reached
limit the strength of the Mpemba effect~\cite{RiosMonje2026}, whereas in
multistep protocols the arrangement of the driving stages affects the overall
time advantage~\cite{PelusoEggerNava2026}. In a nuclear spin network,
polarization and waiting time affect the excitation of slow collective modes
and the crossing time of relaxation curves~\cite{Shah2026}. The comparison
must therefore include both the time spent preparing the state and the time
from release to the chosen tolerance.

Holography offers a tractable framework for studying nonequilibrium dynamics
in strongly coupled systems~\cite{Maldacena1998,GKP1998,Witten1998,Hartnoll2009,LiuSonner2020}.
Quasinormal modes describe relaxation near equilibrium~\cite{HorowitzHubeny2000,KovtunStarinets2005},
and boundary sources applied for a finite time drive the system away from
equilibrium~\cite{LiTianZhang2013,IshiiMurata2018}. After the source is removed,
the bulk fields continue to evolve, and their absorption by the horizon
accounts for dissipation. Following this evolution from the onset of driving
through late relaxation connects the full nonlinear dynamics to the mode
spectrum near equilibrium.

Holographic Mpemba relaxation has been demonstrated in the probe limit~\cite{Ge2026}.
On a fixed background, boundary drives of finite duration have also been used
to study how preparation affects relaxation at matched final charge density~\cite{Shi2026Preparation}.
Nonlinear Maxwell feedback regenerates a slow tail even when the linear slow
response is canceled. A compensating drive can reduce its amplitude, and the
protocols can still yield different arrival times once the duration of driving
is included. The asymptotic slow amplitude therefore depends on both the
projection at release and the subsequent nonlinear feedback.

With gravitational backreaction, the boundary drive also alters the geometry
and can change the total energy, so different protocols may reach different
equilibrium states. Matching the final
energy and charge isolates the effect of preparation in comparisons with a
common endpoint. The evolving horizon area tracks entropy production both
during driving and after release. For given initial and final states, each
increment of horizon entropy during driving reduces the amount still to be
produced during relaxation, so the two stages share a single entropy budget.

Critical slowing down near the superconducting transition provides a natural
setting for examining how this entropy budget constrains the preparation of
a rapidly relaxing state. As the critical point is approached,
the order parameter relaxes more slowly and can influence the dynamics over much
longer times. Optical pump and terahertz probe experiments on niobium nitride
(NbN) have also found that the time required to suppress superconductivity
increases markedly near a dynamical transition~\cite{NavaAntonio2026}. Slow
order parameter dynamics has been studied in Landau descriptions of the Mpemba
effect and in holographic
quenches~\cite{Holtzman2022,Basu:2011ft,Sonner2015,NatsuumeOkamura2017}.
In holographic systems, quasinormal modes describe linear relaxation near
criticality~\cite{Arean2021Symphony}, while equilibrium properties and linear
response help explain the nonlinear dynamics~\cite{Flory2024}. With
backreaction included, the slow field also affects the charge distribution and
geometry~\cite{DonosKailidis2026}. The horizon area then depends both on the
geometric response at a fixed radius and on the motion of the horizon itself.
This connects critical slow dynamics to horizon entropy and provides a basis
for studying physical constraints on preparation.

In this work, we study boundary driving and subsequent relaxation in a
holographic superconductor with full backreaction. At matched final energy and charge,
drives from a common superconducting state can reverse the ordering of the
remaining horizon entropy after release. Some protocols starting from
superconducting states or from scalar-free Reissner--Nordstr\"om (RN) black
holes also reach a given tolerance sooner when driving and waiting times are
counted (section~\ref{sec:results}). Faster relaxation, however, does not
necessarily imply faster preparation. We therefore turn in section~\ref{sec:critical} to a critical state that
already carries a slow scalar field and exterior charge. Removing the field
requires transporting this charge across the horizon. The area law and
charge balance relate the time required for this process to the small
entropy reserve that remains.

\section{Holographic model and horizon entropy}
\label{sec:model}

We consider a minimal holographic $s$-wave superconductor with full
backreaction~\cite{HHH2008,HHH2008b}. The four-dimensional bulk theory couples
Einstein gravity with a negative cosmological constant to a Maxwell field and
a charged complex scalar, with the bulk action
\begin{equation}
 S=\frac{1}{16\pi G}\int\dd^4x\,\sqrt{-g}
 \left[R-2\Lambda+\mathcal L_{\mathrm m}\right]
\label{eq:action}
\end{equation}
and matter Lagrangian
\begin{equation}
 \mathcal L_{\mathrm m}=-\frac14F_{\mu\nu}F^{\mu\nu}
 -(D_\mu\psi)^*D^\mu\psi-m_\psi^2|\psi|^2.
\label{eq:matterLagrangian}
\end{equation}
Here $G$ is Newton's constant, $\Lambda=-3/L^2$, and $L$ is the AdS radius.
The field strength and covariant derivative are $F=\dd A$ and
$D_\mu=\nabla_\mu-\ii qA_\mu$. We set $L=1$, $m_\psi^2=-2$ and $q=2$.
We use standard quantization, for which the dual scalar operator has
dimension $\Delta=2$.

For homogeneous, isotropic evolution, we use ingoing
Eddington--Finkelstein (EF) coordinates~\cite{Murata2010,Bhaseen2013}
and the metric ansatz
\begin{equation}
 \dd s^2=-f(\tau,r)\,\dd\tau^2+2\,\dd\tau\,\dd r
 +\Phi^2(\tau,r)(\dd x^2+\dd y^2).
\label{eq:metric}
\end{equation}
The gauge field is $A=\alpha(\tau,r)\,\dd\tau$, and the scalar field $\psi$
also depends on $(\tau,r)$. The function $\Phi$ sets the scale of the planar spatial directions $x$ and $y$,
while $\alpha$ is the electric potential.
The metric functions and matter fields are determined together by the full
coupled equations. We fix the remaining radial shift by $\Phi-r\to0$ at the boundary.
The derivative along the outgoing null direction and the covariant scalar
flux are
\begin{equation}
 d_+=\partial_\tau+(f/2)\partial_r,\qquad
 \Pi=(d_+-2\ii\alpha)\psi.
\label{eq:nullDefinitions}
\end{equation}
Primes denote radial derivatives and dots denote derivatives with respect to
$\tau$. Varying the action and substituting the field ansatz gives the radial
system
\begin{align}
 \Phi''+\frac{\Phi}{2}|\psi'|^2&=0,
 \label{eq:radialFocusing}\\
 (\Phi^2\alpha')'-4\Phi^2\Imag(\psi^*\psi')&=0,
 \label{eq:radialGauss}\\
 (\Phi d_+\Phi)'-\Phi^2\left(
 \frac32+\frac{|\psi|^2}{2}-\frac{\alpha'^2}{8}\right)&=0,
 \label{eq:radialArea}\\
 (\Phi\Pi)'+(d_+\Phi)\psi'+\Phi\psi+\ii\Phi\alpha'\psi&=0,
 \label{eq:radialScalar}\\
 f''-\frac{4\Phi'd_+\Phi}{\Phi^2}
       +2\Real(\Pi^*\psi')-\alpha'^2&=0.
 \label{eq:radialLapse}
\end{align}
The remaining Einstein and Maxwell equations give
\begin{align}
 d_+^2\Phi-\frac{f'}2d_+\Phi+\frac{\Phi}{2}|\Pi|^2&=0,
 \label{eq:nullFocusing}\\
 d_+(\Phi^2\alpha')+4\Phi^2\Imag(\psi^*\Pi)&=0.
 \label{eq:maxwellTransport}
\end{align}

We start from a static superconducting solution with a regular horizon and
vanishing scalar source, or from an analytic RN solution without scalar hair.
On each ingoing EF slice we solve the radial constraints using Chebyshev
collocation~\cite{Trefethen2000} and advance the fields with the implicit
Crank--Nicolson scheme~\cite{Bhaseen2013}.

Near the AdS boundary, the scalar, electric potential and metric behave as
\begin{equation*}
\begin{aligned}
 \psi&=\frac{J}{r}+\frac{\psi_2}{r^2}+\cdots,
 &\alpha&=\mu-\frac{Q}{r}+\cdots,\\
 \Phi&=r-\frac{|J|^2}{4r}+\cdots,
 &f&=r^2-\frac{|J|^2}{2}-\frac{m}{r}+\cdots .
\end{aligned}
\end{equation*}
Here $J(\tau)$ is the scalar source, $\psi_2$ is the subleading scalar
coefficient, and $\mu$ is the boundary electric potential. We keep the boundary
metric fixed and require the bulk fields to be regular across the future horizon
in ingoing EF coordinates. The source vanishes outside the driving intervals,
and we use a common reference gauge across all stages of the drive. While active,
the source exchanges energy and charge with the system~\cite{Bianchi2002,Murata2010}.
After source removal, $m$ and $Q$ are conserved. They give the endpoint
energy density $E=m/(8\pi G)$ and charge density $Q/(16\pi G)$, both of which
we match across the protocols in each comparison.

The common normal endpoint in each comparison is a planar RN black hole with
$\Phi=r$ and $\psi=0$,
\begin{equation}
 f=r^2-\frac{m}{r}+\frac{Q^2}{4r^2},\qquad
 \alpha=Q\left(\frac{1}{r_h}-\frac{1}{r}\right).
\label{eq:RN}
\end{equation}
Here $r_h$ is the outer horizon. The gauge $\alpha(r_h)=0$ gives the
chemical potential $\mu=Q/r_h$. The temperature and entropy
density are
\begin{equation*}
 T=\frac{1}{4\pi}\left(3r_h-\frac{Q^2}{4r_h^3}\right),
 \qquad s=\frac{r_h^2}{4G} .
\end{equation*}
The surface gravity is $\kappa=2\pi T$, and the first law takes the form
$\dd m=\kappa\,\dd(r_h^2)+(\mu/2)\dd Q$.
We set $Q_f=1$ throughout. Temperatures and times are measured in units of
$\sqrt{Q_f}$ and $Q_f^{-1/2}$, respectively. At this charge the first superconducting
instability occurs at $T_c\simeq0.123368$ and $m_c\simeq0.732072$.

To characterize the late decay, we use the homogeneous scalar quasinormal
modes (QNMs) of the matched RN endpoint. In the full theory, the scalar stress
tensor and current start at quadratic order in perturbations of the RN
background, where the scalar vanishes. Scalar perturbations thus decouple
from the electric and metric sectors at linear order. On a condensed background, perturbations in the three sectors
generally mix~\cite{Bhaseen2013,Arean2021Symphony}.

Let $f_0$ and $\alpha_0$ denote the RN background functions in
eq.~\eqref{eq:RN}. A homogeneous scalar perturbation of the form
$\delta\psi=e^{-\lambda\tau}U(r)$ obeys the linearized Klein--Gordon equation
\begin{equation}
 f_0U''+\left[f_0'+\frac{2f_0}{r}
 -2(\lambda+\ii q\alpha_0)\right]U'
 -\left[m_\psi^2+\ii q\alpha_0'
 +\frac{2}{r}(\lambda+\ii q\alpha_0)\right]U=0.
\label{eq:scalarQNM}
\end{equation}
We write the complex decay exponent as $\lambda=\Gamma+\ii\Omega$. The corresponding QNM frequency is $\omega=-\ii\lambda=\Omega-\ii\Gamma$.
We impose a vanishing scalar source at the boundary and regularity at
the future horizon in EF coordinates~\cite{SonStarinets2002,AmadoKaminskiLandsteiner2009}.
Stable modes have $\Gamma>0$.
At the superconducting critical point, the slow scalar mode reaches
$\lambda=0$ in the chosen RN gauge. We compute the scalar QNMs as a generalized
eigenvalue problem~\cite{Ammon2019Sound,Ammon2022Pseudo}. Throughout this
work we normalize scalar QNMs to unit boundary response.

The apparent horizon is the future outer marginal surface
$r=r_H(\tau)$. A subscript $H$ denotes evaluation on this moving surface.
For radial derivatives, the prime is taken before evaluation at the horizon.
The horizon conditions and its area and entropy densities are~\cite{Bekenstein1973,Hawking1975}
\begin{equation*}
\begin{gathered}
 (d_+\Phi)_H=0,\qquad \Phi'_H>0,\qquad [(d_+\Phi)']_H>0,\\
 \mathcal A_H=\Phi_H^2,\qquad s_H=\frac{\mathcal A_H}{4G}.
\end{gathered}
\end{equation*}
The covariant scalar flux on the horizon is $\Pi_H=\Pi|_H$.
The focusing relation in eq.~\eqref{eq:nullFocusing}, combined with
the horizon motion~\cite{Hayward1994}, gives
\begin{equation}
 \dot{\mathcal A}_H=K_H|\Pi_H|^2,
 \qquad
 K_H=\frac{\Phi_H^2\Phi'_H}{[(d_+\Phi)']_H}>0.
\label{eq:areaLaw}
\end{equation}
The electric field affects area growth through the covariant scalar flux and
the geometry. For the homogeneous radial Maxwell field, its own outgoing
null energy flux vanishes.

For a trajectory approaching the chosen RN endpoint, let
$\mathcal A_f=r_{h,f}^2$ and $s_f=\mathcal A_f/(4G)$.
Integrating the area law gives the normalized remaining entropy,
\begin{equation}
 D(\tau)=\frac{s_f-s_H(\tau)}{s_f}
 =1-\frac{\mathcal A_H(\tau)}{\mathcal A_f}
 =\frac{1}{\mathcal A_f}\int_\tau^\infty
 K_H(\tau')|\Pi_H(\tau')|^2\,\dd\tau' .
\label{eq:area}
\end{equation}
Thus $D(\tau)$ measures the normalized horizon entropy still to be produced as
the black hole approaches its final equilibrium. On the ingoing EF slicing,
the growth of apparent horizon entropy tracks dissipation. When the marginal
surface satisfies the minimar conditions, its horizon entropy equals the
outer entropy~\cite{EngelhardtWall2019}. Because Mpemba crossings can depend
on the measure used to compare relaxation~\cite{Biswas2023dist}, we use
$D(\tau)$ throughout.

After release, the coarse-grained free-energy combination $E_f-T_fs_H$
exceeds its final value by $T_fs_fD$. It therefore gives the same relaxation
ordering as $D$ for trajectories with a common endpoint.
In the probe limit, the corresponding balance describes energy
dissipated into a fixed bath~\cite{Tian2023,Shi2026Preparation}.
For a given initial state and endpoint, horizon growth during driving uses part
of the entropy available for subsequent relaxation.

\section{Mpemba relaxation under boundary driving}
\label{sec:results}

We compare preparations by boundary driving at matched final energy and
charge, first examining entropy reversal from a superconducting state and then
arrival from distinct RN seeds without scalar hair.

\subsection{Entropy reversal and total arrival time}
\label{sec:loop}

We first compare two protocols that take the same superconducting state
to a common RN endpoint at matched energy and charge. Both use smooth complex
scalar sources of finite duration. The direct protocol uses a single pulse.
In the charge loop, the carrier frequency is adjusted to change the phase of
the source relative to the scalar response. This relative phase controls the
direction of charge transfer, allowing the drive to discharge the system and
then recharge it after a period of free evolution. We carry the bulk fields
and their relative phases between stages in a common reference gauge.

\begin{figure}[!t]
 \centering
 \manuscriptpanel{.165}{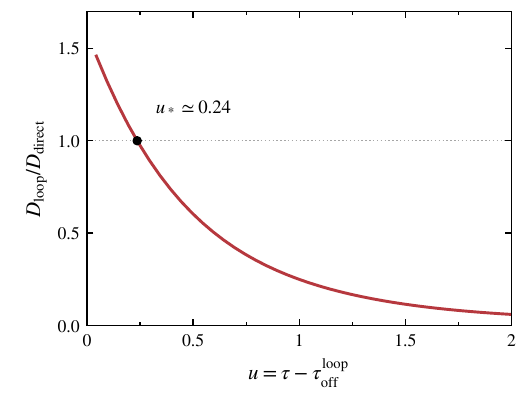}{fig:loop-a}%
 \manuscriptpanel{.155}{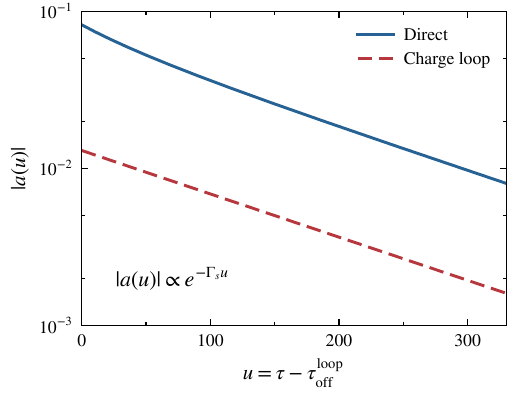}{fig:loop-b}
 \caption{Entropy reversal and amplitudes of the slow mode for drives from $T_i/T_c=0.75$ to $T_f/T_c=1.02$, with $Q_f=1$. Both protocols are shown on a common time axis $u=\tau-\tau_{\rm off}^{\rm loop}$, measured from the end of the loop drive. (a) Remaining entropy of the loop relative to the direct protocol. (b) Magnitude of the projection $a$ onto the slowest scalar QNM.}
 \label{fig:loop}
\end{figure}

After both sources have been removed, we compare the remaining entropy at equal absolute times. The direct source ends earlier, so its state has already relaxed freely for longer. The loop has more remaining entropy when its source ends, but its deficit falls below the direct value about $0.24$ time units after the loop source is removed (figure~\ref{fig:loop-a}). At late times, the loop reduces the slow mode amplitude $|a|$ to about $0.20$ of the direct value. Since both states share the same late decay rate, their entropy ratio approaches the squared amplitude ratio, about $0.040$ (figure~\ref{fig:loop-b}). The discharge and recharge form a Pontus--Mpemba excursion~\cite{NavaEgger2025,Aditya2026}, here with both conserved densities matched.

The entropy reversal raises the question of whether the loop also arrives
earlier once preparation time is included. We vary the discharge plateau
length $T_d$ and the subsequent waiting time $T_w$. For each pair, we adjust
the duration and carrier frequency of the recharge stage to match the
endpoint.

We compare the total arrival times through the ratio
\begin{equation}
 \mathcal R_\delta(T_d,T_w)=
 \frac{\tau_\delta^{\rm dir}-\tau_{\rm start}}
 {\tau_\delta(T_d,T_w)-\tau_{\rm start}} .
 \label{eq:controlRatio}
\end{equation}
The common clock starts at the earliest source onset among the compared
protocols. The time $\tau_\delta$ is the first entry into $D\leq\delta$ at or
after source removal for which the threshold remains satisfied over the
confirmation window.\footnote{We require the threshold to hold at every
sampled time over the following 50 time units.} The total elapsed time
includes driving and waiting up to $\tau_\delta$.

\begin{figure}[!hb]
 \centering
 \manuscriptpanel[.487]{.13656}{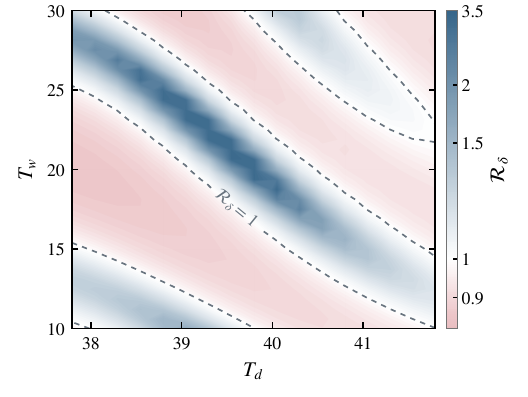}{fig:control}\hfill
 \manuscriptpanel[.487]{.128}{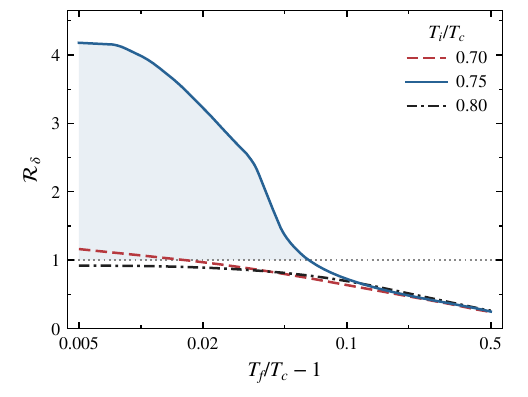}{fig:endpointScan}
 \caption{Total arrival time ratio $\mathcal R_\delta$, defined in eq.~\eqref{eq:controlRatio}, at $\delta=10^{-5}$.
 Values above one favor the charge loop. (a) Dependence on the discharge plateau duration $T_d$ and waiting time $T_w$ for the seed and endpoint of figure~\ref{fig:loop}. The dashed contours separate earlier loop arrival (blue) from later arrival (red).
 (b) Dependence on the endpoint temperature $T_f$ for three seed temperatures $T_i$, with $T_d$ and $T_w$ fixed and $Q_i=Q_f=1$. The shaded region marks earlier loop arrival for $T_i/T_c=0.75$.}
 \label{fig:arrival-scans}
\end{figure}

A ratio above one means that the loop reaches the threshold earlier, even
when driving and waiting are included. Varying the discharge plateau and the
waiting time reveals alternating bands where the loop arrives earlier or
later than the direct protocol (figure~\ref{fig:control}). The largest sampled
ratio in this scan is $3.24$ at $\delta=10^{-5}$.

Figure~\ref{fig:endpointScan} shows how the time advantage varies with the seed
and final temperature at $\delta=10^{-5}$ when the discharge and waiting stages
are held fixed. For $T_i/T_c=0.70$ and $0.75$, the loop arrives earlier at the
colder endpoints, whereas for $T_i/T_c=0.80$ it arrives later throughout the
scan. Close to $T_c$, the slow decay rate $\Gamma_s$ is small. When the slow
tail controls arrival, a given fractional reduction in its amplitude saves
more relaxation time at smaller $\Gamma_s$. The loop can therefore more
readily recover its extra driving and waiting time near $T_c$ if it
suppresses the slow mode.

\subsection{Fast arrival from RN seeds}
\label{sec:near}

We compare two stable RN seeds without scalar hair. Both drives reach the
same endpoint near criticality, with $T_f/T_c=1.001$ and $Q_f=1$.
For each given seed, the clock starts at the onset of driving and excludes
the time spent preparing the seed.

The short protocol uses a compact pulse of duration $T_p=8$ and the baseline
uses two pulses over $T_p=72$. Both target the same slow pole
$\lambda_{\rm ref}$ of the final RN background in the gauge of vanishing
horizon potential. In the continuous linear problem, cancellation requires
the principal source profile $J^{(0)}$ to satisfy
\begin{equation*}
 \int_0^{T_p}e^{\lambda_{\rm ref}\tau}J^{(0)}(\tau)\dd\tau=0.
\end{equation*}
The short drive meets this condition by combining a compact profile with
its time derivative. We tune a small complex correction together with the RN
seed parameters to match the final energy and charge and suppress the slow
projection at release. For the baseline, the relative complex weight of
the two pulses gives the same linear cancellation.

\begin{figure}[!t]
 \centering
 \includegraphics[width=.58\linewidth]{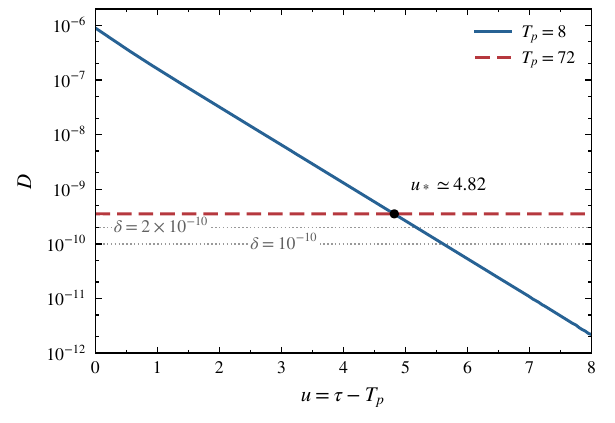}
 \caption{Remaining horizon entropy after release from two distinct RN seeds with the same endpoint, $T_f/T_c=1.001$ and $Q_f=1$. Each curve starts its clock at the end of its own drive, $u=\tau-T_p$. The dot marks the crossing. Dotted lines indicate the two arrival tolerances.}
 \label{fig:near}
\end{figure}

The shorter drive starts from a colder RN seed with a larger entropy
deficit, yet its remaining entropy falls below that of the baseline near $u=4.82$,
where $u=\tau-T_p$ is the time since each drive ended
(figure~\ref{fig:near}). The two drives have the same final spectrum, but
the baseline leaves a larger slow tail. In that tail the scalar amplitude
decays at rate $\Gamma_s\simeq3.09\times10^{-4}$ and the remaining entropy
falls as $e^{-2\Gamma_s u}$. The baseline reaches both tolerances in this
regime. The short protocol reaches them during the fast transient, whose
dominant scalar mode decays at $\Gamma_1\simeq0.801$.

Including the drive, the total time to reach $\delta=2\times10^{-10}$ is
about $13.2$ for the short protocol and $1.01\times10^3$ for the baseline. At
$\delta=10^{-10}$, the total arrival times are about $13.6$ and
$2.13\times10^3$, respectively.\footnote{We apply the confirmation window of
section~\ref{sec:loop} to $|D|\leq\delta$ to allow for small numerical
offsets. The entropy offset from endpoint mismatch is well below both
tolerances. Refined trajectories cover the full confirmation window for the
short protocol at both tolerances and for the baseline at $2\times10^{-10}$.
For the baseline at $10^{-10}$, the time refinement check uses local
restarts and tests of sensitivity to the slow amplitude.}

\subsection{Nonlinear regeneration of the slow tail}
\label{sec:Cs}

Although the short protocol in section~\ref{sec:near} nearly removes the slow
projection at release, nonlinear evolution regenerates a slow tail. The scalar
stress and current drive quadratic changes in the geometry and electric field.
These changes feed back on the scalar at cubic order and excite the slow mode.
Nonlinear excitation also occurs in other black hole channels~\cite{Okuzumi2008,MaYang2024,Sberna2022}.

The amplitude $a(u)$ of the slow mode changes during the subsequent evolution. In the RN gauge used for the mode analysis, its late
tail is $a(u)\sim C_se^{-\lambda_su}$. The coefficient $C_s$ contains both
the slow projection at release and the contribution regenerated by the fast
field. These contributions add as complex amplitudes and can interfere. The
radial reconstruction and projection of the cubic feedback are given in
appendix~\ref{app:compactProjection}.

For the short pulse of duration $T_p=8$, matching leaves an instantaneous
projection $a(0)$ of order $10^{-13}$, whereas nonlinear feedback produces
$C_s$ of order $10^{-10}$. The complete cubic prediction agrees with the
complex coefficient extracted from numerical evolution. Several radial
components of the early fast field contribute through their mixed couplings,
even when the visible transient is dominated by the first fast mode.

\begin{figure}[!hb]
 \centering
 \includegraphics[width=.58\linewidth]{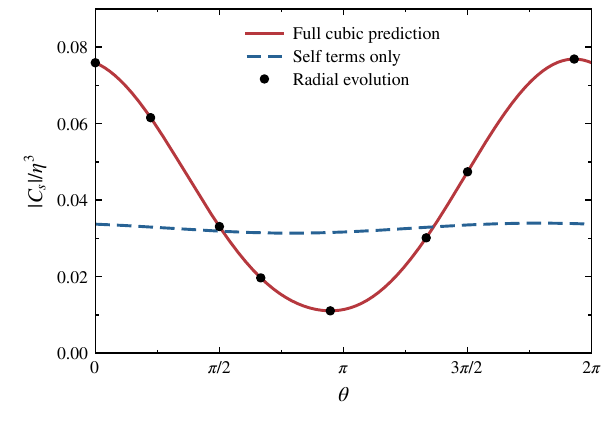}
 \caption{Regenerated slow amplitude as a function of the relative phase $\theta$ of two fast modes. The release amplitudes are $b_1=\eta$ and $b_2=0.75\eta e^{\ii\theta}$, with $\eta=0.008$. Points show the full radial evolution. The solid curve includes all six cubic terms, and the dashed curve retains only the two self terms.}
 \label{fig:regeneration}
\end{figure}

A separate release with two modes isolates the role of mixed couplings. Its two
modes are the first two fast scalar QNMs of the RN background at
$T_f/T_c=1.001$ and $Q_f=1$, ordered by decay rate. Their relative phase is varied at fixed amplitudes
(figure~\ref{fig:regeneration}).\footnote{For this phase scan, we evolve
the fields after release using an implicit backward differentiation formula
(BDF).} The instantaneous slow projection is zero,
and the remaining fields are reconstructed at the same $m_f,Q_f$ to satisfy
the constraints. The complete cubic prediction captures the resulting phase
dependence, while the self terms alone miss much of it. Mixed couplings
therefore play an important role in determining the regenerated tail.

Whether the regenerated tail controls arrival depends on its contribution to the remaining entropy. In the RN gauge with vanishing potential at the horizon, the horizon value of the normalized slow mode is $H_s=U_s(r_{h,f})$. It gives $\psi_H=H_sC_se^{-\lambda_su}+\cdots$, so eq.~\eqref{eq:area} yields
\begin{equation}
 D_s(u)=K_s|C_s|^2e^{-2\Gamma_su}+\cdots,\quad
 K_s=\frac{|\lambda_sH_s|^2}{2\kappa_f\Gamma_s}.
 \label{eq:Ks}
\end{equation}
For a common endpoint and mode normalization, the decay rates and $K_s$ are fixed, while preparation changes $C_s$. The field decays at $\Gamma_s$ and its quadratic entropy contribution at $2\Gamma_s$. The entropy ratio in figure~\ref{fig:loop} therefore follows the squared amplitude ratio. For the short RN release, eq.~\eqref{eq:Ks} gives an estimated prefactor
$K_s|C_s|^2\simeq3.5\times10^{-22}$ for the slow tail contribution to the
remaining entropy, well below the reported tolerances. The fast transient
therefore sets the arrival time at these tolerances, while a much tighter
threshold can probe the residual slow tail.

\section{Critical entropy and preparation constraints}
\label{sec:critical}

Weak driving of a scalar-free RN seed changes its energy and charge at
second order in the drive amplitude. For a stable nonextremal seed of
chemical potential $\mu_i$, the first law gives a quadratic area reserve
if $\delta m-(\mu_i/2)\delta Q$ has a positive coefficient at this order.
This reserve can accommodate a rapidly decaying scalar field whose amplitude
is a sufficiently small multiple of the drive amplitude. The estimate applies
to the normal endpoint in figure~\ref{fig:near} and remains regular as the
endpoint approaches criticality.

On the critical slow branch, the initial state already carries a scalar
field and exterior charge. How quickly can a drive remove them when
preparation and subsequent relaxation draw on the same entropy reserve?
We first determine the size of that reserve from the motion of the horizon.

\subsection{Quartic entropy reserve}

On the critical slow (center) branch at fixed energy and charge, the
exterior geometry and electric field respond quadratically to the scalar
amplitude. The horizon also moves, so its area depends on both the
geometric response and the horizon displacement. The slow mode becomes a
zero mode at criticality, where the constrained nonlinear evolution gives
the equation for its amplitude $a$ and the associated horizon flux
\begin{equation}
\begin{aligned}
 \dot a&=-(g+\ii h)a|a|^2+O(|a|^5),\\
 \Pi_H&=-(g+\ii c)H_ca|a|^2+O(|a|^5).
\end{aligned}
 \label{eq:center}
\end{equation}
The cubic amplitude flow has the structure of the effective dynamics in the
probe limit~\cite{Flory2024}. Here $g>0$, $H_c$ is the horizon value of the
critical mode, and $h$ describes the phase rotation in the chosen gauge. The
linear covariant horizon flux of the critical zero mode vanishes, so the
leading flux is generated by nonlinear feedback. The coefficient $c$ includes the phase rotation and the quadratic potential
shift on the displaced horizon. The coefficients include both geometric and
electric feedback (appendix~\ref{app:compactProjection}).

For the area expansion, set $x=|a|^2$ and $\mathcal A_c=r_{h,c}^2$.
The geometric response, horizon displacement, and area law give
\begin{equation}
\begin{gathered}
 \Phi=r+\Phi_2(r)x+\cdots,\qquad
 r_H=r_{h,c}+r_{H,2}x+\cdots,\\
 \mathcal A_H=\mathcal A_c+
 2r_{h,c}[\Phi_2(r_{h,c})+r_{H,2}]x+O(x^2),\\
 -\dot D=\frac{|H_c|^2(g^2+c^2)}{\kappa_c}x^3+O(x^4).
\end{gathered}
 \label{eq:areaQuadratic}
\end{equation}
Since $\dot x=-2gx^2+O(x^3)$, a nonzero quadratic area coefficient would
give an area derivative of order $x^2$, whereas the flux law requires order
$x^3$. The displacement of the apparent horizon therefore cancels the
quadratic geometric contribution to the area, although the exterior fields
retain their quadratic response. Integrating the area growth along the
critical branch gives
\begin{equation}
 D=K_4|a|^4+O(|a|^6),\qquad
 K_4=\frac{|H_c|^2(g^2+c^2)}{4\kappa_cg}.
\label{eq:K4}
\end{equation}
Figure~\ref{fig:crossover} shows this quartic dependence on the scalar
amplitude. Along the source-free critical flow, the amplitude equation gives
$|a|\propto\tau^{-1/2}$, and hence $D\propto\tau^{-2}$.

\subsection{A common entropy coefficient for equilibrium and relaxation}
\label{sec:crossover}

Away from criticality, the area response acquires a term quadratic in the
scalar amplitude. Its coefficient can be determined from the nearby
superconducting equilibrium. For each nearby RN reference at $Q=1$, define
$\delta_T=T_{\rm RN}(m,1)/T_c-1$.

Near criticality, the slow pole and amplitude flow take the form
\begin{equation*}
\begin{aligned}
 \lambda_s&=(\Gamma_T+\ii\Omega_T)\delta_T+O(\delta_T^2),\\
 \dot x&=-2x(\Gamma_s+gx)+\cdots,\\
 x_{\rm eq}&=-\Gamma_T\delta_T/g+\cdots .
\end{aligned}
\end{equation*}
The last relation gives the equilibrium amplitude on the hairy side, where
linear response and nonlinear damping balance.

At fixed $m,Q$, the local slow branch is parametrized by the scalar amplitude.
The field equations fix the fast response without independently exciting fast
modes. Let $\mathcal A(\delta_T,x)$ be the area on this branch and
$\mathcal A_{\rm RN}(\delta_T)$ the RN area at the same $m,Q$.
We define a signed area deficit relative to RN and expand it on this branch.
The remaining entropy $D$ of eq.~\eqref{eq:area} is instead measured relative
to the final equilibrium of a relaxation process.

The critical expansion fixes the coefficient of $x^2$ to $K_4$. The nearby
hairy equilibrium belongs to the continuation of this branch to the order
considered. Near it, an amplitude perturbation relaxes linearly, while the
squared flux and area production are quadratic in the perturbation. The area therefore
has no variation to first order at $x_{\rm eq}$ along the branch at fixed $m,Q$.
This stationarity fixes the mixed coefficient of $\delta_Tx$ and gives
\begin{equation}
 1-\frac{\mathcal A(\delta_T,x)}{\mathcal A_{\rm RN}(\delta_T)}
 =\frac{2K_4\Gamma_T}{g}\delta_Tx+K_4x^2
 +O(\delta_T^2x+\delta_Tx^2+x^3).
 \label{eq:signedArea}
\end{equation}
This deficit can be negative on the hairy side, where the equilibrium area
exceeds the RN area.
Its equilibrium value and the quadratic weight on the normal side as
$\Gamma_s\to0^+$ yield
\begin{align}
 \frac{\mathcal A_{\rm hair}-\mathcal A_{\rm RN}}{\mathcal A_c}
 &=K_4x_{\rm eq}^2+O(x_{\rm eq}^3),\label{eq:equilibriumArea}\\
 K_s&=\frac{2K_4}{g}\Gamma_s+O(\Gamma_s^2).
 \label{eq:normalEntropyWeight}
\end{align}
The same coefficient $K_4$ thus connects the critical entropy reserve, the
nearby equilibrium entropy advantage and the entropy weight of the slow tail
on the normal side. These relations follow from a single constrained response
of the horizon area. The hairy equilibrium has the larger area at fixed energy and charge.
It is a local entropy maximum along the slow branch, consistent with the
stability of its homogeneous amplitude.

\begin{figure}[!t]
 \centering
 \includegraphics[width=.58\linewidth]{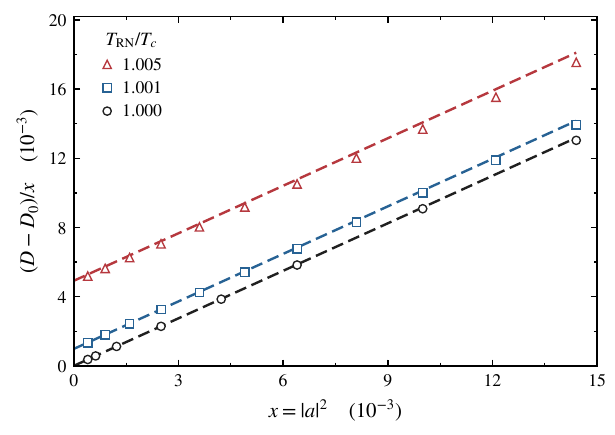}
 \caption{Critical entropy reserve and crossover on the normal side. Open symbols give the horizon area response of the constrained slow states at three RN reference temperatures. Here $x=|a|^2$ and $D_0$ is the reconstructed RN deficit at $x=0$. In $(D-D_0)/x$, the quartic coefficient $K_4$ gives the leading slope near criticality. The intercept vanishes at the critical point and approaches the entropy weight $K_s$ of the slow tail on the normal side. Dashed lines give the predictions of eqs.~\eqref{eq:signedArea}--\eqref{eq:normalEntropyWeight} with independently calculated coefficients.}
 \label{fig:crossover}
\end{figure}

In the RN gauge, the equilibrium scalar can rotate while its covariant
horizon flux vanishes. This condition relates the phase response to damping,
with $\Omega_T/\Gamma_T=c/g$ at leading order.
With this relation and the critical values of the horizon quantities,
eq.~\eqref{eq:Ks} reproduces eq.~\eqref{eq:normalEntropyWeight}. At fixed
slow amplitude, the squared flux scales as $\Gamma_s^2$, while the time
integration of the exponential decay contributes a factor proportional to
$\Gamma_s^{-1}$. Hence $K_s\propto\Gamma_s$, and a slow field can persist for
a long time while contributing little to the remaining entropy.

The first law at fixed charge provides an independent static estimate of
$K_4$ from the nearby hairy branch. Its derivation and comparison with the
horizon flux result are given in appendix~\ref{app:compactProjection}.

Solving the radial constraints along the local slow family at fixed $m$ and
$Q$ and locating the apparent horizon gives the area samples in
figure~\ref{fig:crossover}. The samples follow the leading scaling near
criticality, with larger deviations as the amplitude increases and the
temperature moves away from its critical value. The crossover at
$gx\sim2\Gamma_s$ separates the quadratic regime at smaller amplitudes from
the quartic regime at larger amplitudes. At any fixed
endpoint on the normal side, sufficiently small amplitudes eventually enter the
quadratic regime. At strict criticality the quartic term remains leading. The
instantaneous slow amplitude approaches $C_se^{-\lambda_su}$ only when
$g|a|^2\ll\Gamma_s$.
Near criticality, the endpoint and the chosen entropy tolerance therefore
both affect which part of the slow relaxation is resolved.

\subsection{Entropy cost of a fast release}
\label{sec:budget}

The quartic reserve constrains the state left at release. We start from late
states on the critical slow branch with common $m_c,Q_c$ and initial horizon fields of
order $\varepsilon=|a_i|$ (appendix~\ref{app:compactChargeBound}). Energy
and charge can vary during driving and return to their initial values at
release. The subsequent relaxation to critical RN shares the initial entropy
reserve with preparation,
\begin{align}
 D_i&=K_4\varepsilon^4+O(\varepsilon^6),
 \label{eq:initialCriticalReserve}\\
 D_i&=\frac{\mathcal A_{\rm off}-\mathcal A_i}{\mathcal A_c}
               +D_{\rm off}.
 \label{eq:budgetPartition}
\end{align}
Here $D_i$ is the initial remaining entropy and
$D_{\rm off}=1-\mathcal A_{\rm off}/\mathcal A_c$ is its value at release.
Both contributions are nonnegative along a future outer apparent horizon
approaching the endpoint.

We consider release states that relax exponentially after the drive is
removed, with no algebraic tail from the critical mode. We assume these
states lie on a local strong stable manifold near critical RN, where the
fast amplitudes determine the nonlinear slow component~\cite{BatesJones1989}.
Appendix~\ref{app:compactChargeBound} specifies the estimates that control
the horizon field and the reconstruction of the constraints. Consider a
fixed finite set of modes with distinct decay exponents
$\lambda_j=\Gamma_j+\ii\Omega_j$, $\Gamma_j>0$, and nonzero horizon values
$H_j$. For release amplitudes $b_j$, the future horizon flux gives
\begin{equation}
 D_{\rm off}=\frac1{\kappa_c}\int_0^\infty
 \left|\sum_j\lambda_jH_jb_je^{-\lambda_ju}\right|^2\dd u+\cdots .
 \label{eq:releaseCost}
\end{equation}
The omitted terms are higher order in the release amplitudes
(appendix~\ref{app:compactChargeBound}). Distinct exponents and nonzero
horizon values make the quadratic form positive definite. For sufficiently
small releases in this fixed family, the entropy cost is therefore at least
a fixed positive multiple of $\sum_j|b_j|^2$, uniformly as $\varepsilon\to0$.
Since $D_{\rm off}\leq D_i$ and the initial reserve is quartic in $\varepsilon$,
the mode amplitudes and the horizon field at release satisfy
\begin{equation}
 b_j=O(\varepsilon^2),\qquad
 \psi_{H,\rm off}=O(\varepsilon^2).
 \label{eq:releaseField}
\end{equation}
Reaching such a release requires changing the horizon field by
$O(\varepsilon)$ within an entropy budget of $O(\varepsilon^4)$. The
quadratic dependence of area growth on the covariant flux makes the entropy
cost depend on the duration of preparation, while charge balance constrains
the path traced by the integrated flux.

\subsection{Preparation time from entropy and charge transport}
\label{sec:chargeTransport}

Completing the preparation also transfers the exterior charge carried by
the initial scalar. The ratio $R=c/g>0$ compares the covariant phase response
with amplitude damping. At the common critical charge $Q_c=Q_f=1$, the
horizon and exterior charges on an EF slice satisfy
\begin{align}
 q_H&=\Phi_H^2\alpha'_H,\qquad Q_{\rm ext}=Q-q_H,\notag\\
 Q_{{\rm ext},i}&=2\mathcal A_cR|\psi_{H,i}|^2+O(\varepsilon^4).
 \label{eq:chargeRatio}
\end{align}
The last relation follows from the Maxwell balance and critical amplitude
flow (appendix~\ref{app:compactProjection}) and gives the exterior charge
per unit horizon scalar intensity. It is
independent of the critical mode's normalization and is determined by the
radial charge distribution of the critical zero mode
(appendix~\ref{app:compactProjection}).
For the finite family above, release at the same total charge has
$Q_{{\rm ext},{\rm off}}=O(\varepsilon^4)$.
Charge of order $\varepsilon^2$ must therefore cross the horizon while
area growth is limited to order $\varepsilon^4$.

The area and Maxwell balances on the horizon give a preparation time bound
in terms of the field and charge at release without a mode decomposition.
The critical initial states in section~\ref{sec:budget} have common values
$m_c,Q_c$ and obey the uniform expansions in appendix~\ref{app:compactChargeBound}.
We allow energy and charge to vary during preparation and require the drive
to restore both values at release. We assume a future outer apparent
horizon that remains smooth throughout preparation and approaches the common
nonextremal critical RN endpoint. When the radial ratio
$|(D_r\psi)_H|/(2\Phi_H\Phi'_H)$ stays uniformly bounded over preparation as
$\varepsilon\to0$, total area growth keeps the accumulated radial drift
subleading to the required change in the horizon field. The physical
release conditions are
\begin{equation}
 \psi_{H,\rm off}=o(\varepsilon),\qquad
 q_{H,\rm off}=Q_c+o(\varepsilon^2).
 \label{eq:physicalReleaseConditions}
\end{equation}
The releases in the fixed finite family of section~\ref{sec:budget} satisfy
eq.~\eqref{eq:physicalReleaseConditions}.

\begin{figure}[!t]
 \centering
 \includegraphics[width=.50\linewidth]{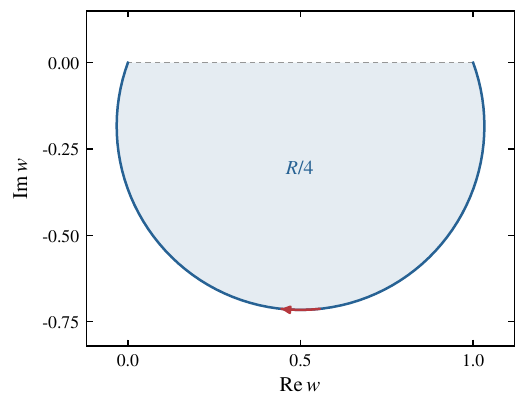}
 \caption{Minimum action for the path defined by the integrated covariant flux. The blue arc connects $w(0)=1$ to $w(1)=0$ and gives the minimum flux action $G(R)$ of eq.~\eqref{eq:chargeAction} for the critical ratio $R=c/g$. Charge transport fixes the shaded area enclosed by the arc and the segment on the real axis to $R/4$. The arrow indicates the orientation.}
 \label{fig:charge_path}
\end{figure}

Let $\mathcal T$ be the full preparation duration, including
waiting. We set $\tau=0$ at preparation onset and use normalized time
$s=\tau/\mathcal T$. Starting from $w(0)=1$, we obtain the complex path $w(s)$ by integrating
the parallel transported covariant flux and normalizing the increment by the
initial horizon field (appendix~\ref{app:compactChargeBound}). Under the same radial control, it
agrees at leading order with the horizon field after parallel transport,
divided by its initial value. To this order the path connects $1$ to $0$. The minimum flux action on the unit interval
and the charge constraint are
\begin{equation}
 \begin{gathered}
 G(R)=\min\int_0^1|\partial_s w|^2\dd s,\\
 \int_0^1\Imag(\bar w \partial_s w)\dd s=-R/2.
 \end{gathered}
 \label{eq:fluxActionConstraint}
\end{equation}
Eq.~\eqref{eq:fluxActionConstraint} gives an isoperimetric problem in
the complex $w$ plane. Closing the path along the real axis makes the charge
integral twice the oriented area of the loop, so Maxwell balance fixes this
area to $-R/4$ (figure~\ref{fig:charge_path}). The area law bounds the kinetic
action of the path, which the straight line at constant speed would minimize
if only the endpoints were fixed. The required charge transfer forces the
path to bend and raises the minimum flux cost.

These endpoint and charge constraints follow from the full evolution in the
critical limit. A boundary drive selects a particular path through the coupled
bulk dynamics. Minimizing over all paths with these constraints gives a lower
bound on its flux cost. The minimum is attained by a circular arc. With $\theta$ the angle
swept by the arc, $0<\theta<2\pi$,
\begin{equation}
 R=\frac{\theta-\sin\theta}{1-\cos\theta},\qquad
 G(R)=\frac{\theta^2}{2(1-\cos\theta)}.
\label{eq:chargeAction}
\end{equation}
The straight path at constant speed has unit action, $G(0)=1$, and zero
oriented area. Charge transfer raises the minimum to $G(R)>1$. For the present
critical mode, $R\simeq2.547$ and $G(R)\simeq4.191$.

The fraction of the initial entropy reserve left at release is
$\eta_{{\rm rem},\varepsilon}=D_{\rm off}/D_i$. The drive uses the
complementary fraction $1-\eta_{{\rm rem},\varepsilon}$ as the horizon area
increases by $\Delta\mathcal A=\mathcal A_{\rm off}-\mathcal A_i$.
Along sequences on which $\mathcal T\Delta\mathcal A/\varepsilon^2$
approaches its finite lower limit as $\varepsilon\to0$, the horizon balances in
appendix~\ref{app:compactChargeBound} imply
\begin{equation*}
 \mathcal T\Delta\mathcal A\geq
 \frac{\mathcal A_c}{\kappa_c}|\psi_{H,i}|^2[G(R)-o(1)].
\end{equation*}
The squared initial field on the right is of order $\varepsilon^2$.
Combining this estimate with the quartic entropy budget in
eqs.~\eqref{eq:initialCriticalReserve}
and~\eqref{eq:budgetPartition} gives the preparation time bound
\begin{equation}
 \liminf_{\varepsilon\to0}
 \varepsilon^2\mathcal T(1-\eta_{{\rm rem},\varepsilon})
 \geq\frac{4gG(R)}{g^2+c^2}.
 \label{eq:chargeTime}
\end{equation}
Even with the entire entropy reserve available for preparation, the time is
at least of order $\varepsilon^{-2}$. Leaving more entropy for later relaxation
raises this lower bound. If $\eta_{{\rm rem},\varepsilon}\to1$, then
$\varepsilon^2\mathcal T$ must diverge.
Along the family of critical initial states, the lower bound on
$\mathcal T(1-\eta_{{\rm rem},\varepsilon})$ also grows as
$D_i^{-1/2}$ or $Q_{{\rm ext},i}^{-1}$. As the initial state approaches
critical equilibrium, the exterior charge vanishes and the entropy reserve
scales as its square, a relation supported by measurements of the horizon
area and electric flux in source-free critical evolutions.

A useful reference is the time $\mathcal T_{1/2}$ for source-free evolution
to halve the modulus of the horizon field. Natural halving removes a fixed fraction
of the initial field, whereas a release satisfying
eq.~\eqref{eq:physicalReleaseConditions} suppresses its leading amplitude and
exterior charge. The leading halving time follows by integrating
the critical flow in eq.~\eqref{eq:center}. Since $\psi_H=H_ca+O(|a|^3)$, this physical target has the same leading
time as halving $|a|$. When
$\eta_{{\rm rem},\varepsilon}\to\eta_{\rm rem}<1$, the bound gives
\begin{equation}
 \limsup_{\varepsilon\to0}
 \frac{\mathcal T_{1/2}}{\mathcal T}
 \leq(1-\eta_{\rm rem})\frac{3(1+R^2)}{8G(R)}.
 \label{eq:chargeSpeed}
\end{equation}
For this release target, the entropy constraint alone gives an upper bound
of $2.81(1-\eta_{\rm rem})$ on the ratio of the natural halving time to the
preparation time. Charge transport reduces this upper bound by a factor
$G(R)\simeq4.191$, to $0.670(1-\eta_{\rm rem})$. It is therefore below
unity for $0\leq\eta_{\rm rem}<1$. In this asymptotic regime, the prescribed
preparation takes longer than free decay to half the initial horizon field
amplitude.

\section{Conclusions and discussion}
\label{sec:discussion}

In this work, we have studied boundary preparation and the subsequent Mpemba
relaxation in a holographic superconductor with full gravitational
backreaction. For given initial and final states, horizon entropy produced
during driving reduces the entropy remaining for relaxation. Our numerical
results show that boundary drives from the same superconducting state can
reverse the ordering of the remaining horizon entropy after source removal.
A separate comparison at fixed entropy tolerance shows that some protocols
retain their time advantage once driving and waiting are counted. For
distinct scalar-free RN seeds, we also find a net time saving when each
protocol is timed from the onset of driving. Nonlinear electric and
geometric feedback can regenerate
a slow tail after source removal, yet a fast transient can determine arrival
if the tail's contribution to the remaining entropy stays below the target
tolerance. Eliminating the tail entirely and reaching a finite tolerance as
early as possible can therefore be different optimization problems. The
entropy available for preparation also depends on the initial state. Under
weak driving, RN initial states can provide a quadratic entropy budget for
building a scalar field. Removing an existing scalar field and its exterior
charge from a critical slow state instead draws on a quartic entropy reserve
that shrinks rapidly near equilibrium.

To understand why this critical reserve is so small, we have examined the
motion of the apparent horizon. The scalar field changes the electric field
and geometry at quadratic order and also displaces the horizon. Along the
critical slow branch at fixed energy and charge, the contribution of the
horizon displacement to the area cancels the quadratic response at a fixed
radius. The coefficient $K_4$ of the remaining quartic term also determines
the entropy advantage of the nearby superconducting equilibrium and the
leading entropy weight of the linear slow tail on the normal side. As the
critical point is approached from that side, the slow mode's lifetime grows
while its entropy weight vanishes in proportion to its decay rate at fixed
mode normalization. The scalar field can thus persist for a long time while
contributing very little to the remaining entropy. An entropy tolerance and
a tolerance on the order parameter can then identify different stages of
relaxation, illustrating why Mpemba ordering depends on the comparison
measure~\cite{Biswas2023dist}.

The cost of preparing a rapidly relaxing state depends on the controls
available. In an open Markovian quantum system, an initial unitary
transformation can eliminate excitation of the slowest mode, while coherent
rotations combined with nonselective measurements can introduce mixing of the
state~\cite{Carollo2021,Hong2026Steering}. In granular fluids, the range of
reachable velocity distributions limits the strength of the Mpemba effect,
while in multistep driving the arrangement of the stages affects whether the
overall process saves time~\cite{RiosMonje2026,PelusoEggerNava2026}. For
systems coupled to a heat bath, work extraction and nonequilibrium free energy
offer complementary thermodynamic perspectives~\cite{ChetriteKumarBechhoefer2021,Moroder2024}.
In a gravitational system, driving can also change the spacetime background on
which control relies. Near a black hole exceptional point, shaping a boundary
pulse can remove the factor of time multiplying the leading exponential in
linear response while preserving its decay rate. At small but finite pulse
amplitude, the drive heats the black hole and splits the coalesced poles,
shortening the window in which this decay law can be observed~\cite{Shi2026EP}.

The tradeoff between time and dissipation appears in accelerated Brownian
equilibration~\cite{Martinez2016}. Thermodynamic speed limits and optimal
transport relate the duration of a process to entropy production and to
dynamical activity or mobility~\cite{Shiraishi2018,VuSaito2023}. Experiments
have also shown that optimized driving can reduce dissipation~\cite{Yan2022,Shen2026Dissipation}.
Among stochastic models, reversible Markov systems with tunable kinetic
barriers provide a close parallel to our setting because the initial
distribution and final equilibrium fix the entropy budget~\cite{WangLu2026Barriers}.
In the critical preparation studied here, the horizon scalar field must
change by $O(\varepsilon)$, while the available entropy reserve is only
$O(\varepsilon^4)$. Since horizon entropy production is quadratic in the
flux, the conditions in section~\ref{sec:chargeTransport} give a lower bound
of order $\varepsilon^{-2}$ on the preparation time. A driven system coupled
to an external heat bath can dissipate supplied work into its surroundings
even at fixed system endpoints. Extending the argument to such systems
requires identifying the part of the entropy budget that remains constrained
by the available controls.

Charge transport further constrains the same covariant scalar flux that
produces horizon entropy. At leading order, paths connecting the same values
of the horizon scalar field after parallel transport can carry different
amounts of charge. The Maxwell balance relates this charge transfer to the
oriented area enclosed by the normalized integrated flux path and its
endpoint chord. The minimum dissipation therefore depends on both the
endpoint separation and the prescribed charge transfer. Thermodynamic
geometry studies dissipation along paths in linear response~\cite{SivakCrooks2012,Scandi2022},
while adiabatic thermal machines lead to isoperimetric optimization of closed
control cycles~\cite{TerrenAlonso2022}. Here charge conservation imposes an
isoperimetric constraint on an open flux path. Joint measurements of particle
and entropy transport in fermionic superfluids~\cite{Fabritius2024} suggest
a possible setting for exploring how transport affects preparation at matched
final energy and particle number.

At fixed charge transfer, the horizon flux action reaches its minimum $G(R)$
when the normalized integrated flux traces a circular arc at constant speed
in normalized time. Starting from critical initial states of decreasing
amplitude, whether an actual boundary drive satisfying the release conditions
can approach this arc through the full nonlinear evolution remains an open
question.

\acknowledgments
This work was supported by the National Natural Science Foundation of China under Grants No. 12305063 and No. 12547143.
\section*{Data availability}
The numerical data, including the source and initial state inputs at full precision and the associated numerical settings, are available from the corresponding author upon reasonable request.

\appendix

\section{Nonlinear response near criticality}
\label{app:compactProjection}

At fixed $m,Q$, the quadratic response of the geometry and electric field
produces a cubic correction to scalar evolution. Its projection onto the slow
mode determines the nonlinear tail discussed in section~\ref{sec:Cs}. The associated
horizon response gives the critical entropy in eq.~\eqref{eq:K4} and the
exterior charge in eq.~\eqref{eq:chargeRatio}.

For the scalar variable $p=r^2\psi$, source-free evolution takes the form
$\dot p=Lp+\mathcal N(p)$, where
$\mathcal N(p)=\mathcal N_3[p]+O(\|p\|^5)$. For the slow mode
$r_s=r^2U_s$, we choose the left functional so that
\begin{equation}
\begin{gathered}
 Lr_s=-\lambda_sr_s,\quad \ell_sL=-\lambda_s\ell_s,\quad \ell_sr_s=1,\\
 a=\ell_sp,\qquad B=(I-r_s\ell_s)p(0).
\end{gathered}
\label{eq:modeProjection}
\end{equation}
Here $B$ is the fast field at release and $a$ is the slow amplitude used in
the main text~\cite{Warnick2015,EngelNagel2000}. For releases with
$a(0)=O(\|B\|^3)$, the leading nonlinear tail is~\cite{Shi2026Preparation}
\begin{equation}
 C_s\simeq a(0)+\int_0^\infty e^{\lambda_su}
          \ell_s\mathcal N_3[e^{uL}B]\,\dd u.
\label{eq:CsCubic}
\end{equation}
The two terms are complex amplitudes. The first is the release projection and
the second is regenerated by the leading fast evolution. The radial equations
below determine $\mathcal N_3$ including the electric and geometric feedback.

We use the RN gauge
$\alpha_0(r_{h,f})=0$ and fix the radial shift by $\Phi-r\to0$.
\begingroup\widowpenalty=10000
For $\psi=\eta\varphi$, with real $\eta$,
$\varphi=\varphi_2/r^2+\cdots$ and $k_0=f_0/2$, write
\par\endgroup\nopagebreak[3]
\begin{equation}
\begin{aligned}
 (\Phi,\alpha,f)&=(r,\alpha_0,f_0)
       +\eta^2(\Phi_2,\alpha_2,f_2)+\cdots,\\
 (\Phi^2\alpha',\Phi d_+\Phi)&=(Q,rk_0)+\eta^2(E_2,Z_2)+\cdots,\\
 \Phi\Pi&=\eta V_1+\eta^3V_3+\cdots .
\end{aligned}
\label{eq:methodResponseExpansion}
\end{equation}
All fields are regular through the
future horizon. At infinity, $\Phi_2,\Phi_2'\to0$, $\alpha_2\to0$ and
$f_2,f_2'\to0$, while $V_1\to-\varphi_2$ and $V_3\to0$.
Setting $E_2(\infty)=Z_2(\infty)=0$ keeps $Q$ and $m$ fixed.

The scalar stress and current determine the quadratic geometry and electric
field through the radial constraints. The coefficient $E_2$ gives the quadratic
correction to the radial electric flux, while $V_1$ and $V_3$ are scalar flux
coefficients. The quadratic response of $d_+\Phi$ is $P_2=(Z_2-\Phi_2k_0)/r$.
Eqs.~\eqref{eq:radialFocusing}--\eqref{eq:radialArea} give
\begin{align}
 \Phi_2''&=-\frac r2|\varphi'|^2,
 \label{eq:methodQuadraticResponseS}\\
 E_2'&=4r^2\Imag(\varphi^*\varphi'),
 \label{eq:methodQuadraticResponseE}\\
 \alpha_2'&=\frac{E_2}{r^2}-\frac{2Q\Phi_2}{r^3},
 \label{eq:methodQuadraticResponseAlpha}\\
 Z_2'&=3r\Phi_2+\frac{r^2}{2}|\varphi|^2
       -\frac{r\Phi_2\alpha_0'^2+r^2\alpha_0'\alpha_2'}4 .
 \label{eq:methodQuadraticResponseZ}
\end{align}
Eqs.~\eqref{eq:radialScalar} and~\eqref{eq:radialLapse} give the
scalar flux and the remaining metric response,
\begin{align}
 V_1'&=-k_0\varphi'-r\varphi-\ii r \alpha_0'\varphi,
 \label{eq:methodFluxResponseV1}\\
 f_2''&=\frac4{r^2}
       \left(P_2+\Phi_2'k_0-\frac{2\Phi_2k_0}{r}\right)
       -2\Real[(V_1/r)^*\varphi']+2\alpha_0'\alpha_2',
 \label{eq:methodFluxResponseF}\\
 V_3'&=-P_2\varphi'-\Phi_2\varphi
       -\ii(\Phi_2\alpha_0'+r\alpha_2')\varphi .
 \label{eq:methodFluxResponseV3}
\end{align}
\begingroup\interlinepenalty=10000
These quadratic fields change scalar propagation at cubic order.
Substituting them into the scalar equation gives
\par\endgroup\nopagebreak[3]
\begin{equation}
\begin{aligned}
 L_\psi\varphi&=V_1/r-k_0\varphi'+2\ii \alpha_0\varphi,\\
 \cN_{3,\psi}[\varphi]&=\frac{V_3}{r}-\frac{\Phi_2V_1}{r^2}
       -\frac{f_2}{2}\varphi'+2\ii\alpha_2\varphi.
\end{aligned}
\label{eq:methodN3}
\end{equation}
In the $p$ variable, $Lp=r^2L_\psi[p/r^2]$ and
$\cN_3[p]=r^2\cN_{3,\psi}[p/r^2]$.
For a time-dependent profile, the quadratic geometry also evolves. Differentiating
the quadratic constraint along the linear history $\dot\varphi=L_\psi\varphi$
determines $\dot\Phi_2$. This contribution enters the cubic response through
$P_2=\dot\Phi_2+f_2/2+k_0\Phi_2'$.

For the critical zero mode $r_c$, we normalize the left functional by
$\ell_cr_c=1$. The center expansion
$p=ar_c+a|a|^2p_3+\cdots$, with $\ell_cp_3=0$, gives
\begin{equation}
\begin{aligned}
 g+\ii h&=-\ell_c\cN_3[r_c],\\
 L_cp_3&=-(I-r_c\ell_c)\cN_3[r_c].
\end{aligned}
\label{eq:compactProjection}
\end{equation}
Here $L_c$ is the scalar evolution generator on the critical RN background.
The projection fixes the cubic amplitude coefficient, while the complementary
equation determines the profile correction without changing the slow amplitude.
For a general released fast field, the same reconstruction along
$\varphi(u)=r^{-2}e^{uL}B$ supplies the cubic feedback in the slow-mode
projection above.

To obtain the entropy coefficient, we evaluate the flux on the displaced
horizon. The explicit radial displacement term cancels in the scalar flux.
The horizon displacement, potential shift and flux coefficient are
\begin{equation*}
\begin{aligned}
 r_{H,2}&=-P_2(r_{h,c})/\kappa_c,\qquad
 \alpha_{H,2}=\alpha_2(r_{h,c})+r_{H,2}\alpha_0'(r_{h,c}),\\
 c&=h+2\alpha_{H,2}.
\end{aligned}
\end{equation*}
The resulting entropy coefficient $K_4$ is given in eq.~\eqref{eq:K4}.

The equilibrium hairy branch provides a separate static determination of
$K_4$. At fixed $Q$, we use the equilibrium value of
$x=|a|^2$ as the branch parameter. Each RN reference has the same $m(x),Q$.
The first law then gives
\begin{equation}
 \frac{\dd(\mathcal A_{\rm hair}-\mathcal A_{\rm RN})}{\dd x}
 =\frac{\dd m}{\dd x}
 \left(\frac1{\kappa_{\rm hair}}-\frac1{\kappa_{\rm RN}}\right).
 \label{eq:staticFirstLaw}
\end{equation}
The surface gravities are evaluated on their respective equilibria and
coincide at criticality. Integrating the term linear in $x$ on the right gives
the quadratic area coefficient $\mathcal A_cK_4$ in
eq.~\eqref{eq:equilibriumArea}. Finite differences from the common critical
solution give the static estimates in table~\ref{tab:criticalCoefficients}.

On the critical slow branch, write $q_H=Q_c+q_2x+O(x^2)$. With the
amplitude flow $\dot x=-2gx^2+O(x^3)$, the leading Maxwell balance reads
$-2gq_2=4\mathcal A_c|H_c|^2c$. Since
$|\psi_H|^2=|H_c|^2x+O(x^2)$, this relation fixes the initial exterior
charge in eq.~\eqref{eq:chargeRatio}.
For the critical scalar profile $\varphi=r_c/r^2$, the zero-mode equation
and Gauss's law also relate $R=c/g$ to its radial charge distribution,
\begin{equation}
 R=\frac{4}{\mathcal A_c|H_c|^2}\int_{r_{h,c}}^\infty
 \frac{r^2\alpha_0}{f_0}|\varphi|^2\dd r.
 \label{eq:criticalChargeProfile}
\end{equation}
Table~\ref{tab:criticalCoefficients} compares this integral with the ratio
extracted from the cubic response. Both calculations use the same RN gauge. The flux and area expansions give consistent
estimates of $K_4$ that remain stable at the displayed precision as the
amplitude window and expansion order vary. The static estimates use two finite amplitudes without
extrapolation.

\begin{table}[tbp]
 \caption{Independent estimates of the critical coefficients. $N$ is the
 radial spectral order. Entries at $N=40,60$ agree at the displayed precision.
 The two static values correspond to squared boundary amplitudes
 $3.97\times10^{-7}$ and $4.63\times10^{-5}$ at $Q=1$. A dash marks a coefficient not evaluated by that method.}
 \label{tab:criticalCoefficients}
 \centering
 \small
 \setlength{\tabcolsep}{4pt}
 \begin{tabular}{@{}lccc@{}}
  \toprule
  Calculation & $N$ & $K_4$ & $R$ \\
  \midrule
  Cubic response & $40,60$ & $0.91583$ & $2.54697$ \\
  Area expansion & $40,60$ & $0.91583$ & \textemdash \\
  Static first law & $80$ & $0.9158,\ 0.9160$ & \textemdash \\
  Radial charge integral & $40,60$ & \textemdash & $2.54697$ \\
  \bottomrule
 \end{tabular}
\end{table}

\section{Entropy constraints on fast preparation}
\label{app:compactChargeBound}

We derive the time bound under the horizon, radial control and release
conditions of section~\ref{sec:chargeTransport}. The initial horizon field obeys
$|\psi_{H,i}|^2=|H_c|^2\varepsilon^2+O(\varepsilon^4)$ uniformly along the
critical family. Let $M_r$ denote the supremum of
$|(D_r\psi)_H|/(2\Phi_H\Phi'_H)$ over the full preparation
interval. This supremum is bounded uniformly as $\varepsilon\to0$.
The area increase during preparation is
$\Delta\mathcal A=\mathcal A_cD_i(1-\eta_{{\rm rem},\varepsilon})=O(\varepsilon^4)$.
Monotonicity gives $\mathcal A_i\leq\mathcal A_H\leq\mathcal A_c$.

\paragraph{Fast releases.}
Let $B=(I-r_c\ell_c)p$ denote the component complementary to the critical
slow mode. We consider a local strong stable graph on which
$a_{\rm off}=O(\|B_{\rm off}\|^3)$, with a decay gap and evolution estimates
controlling the horizon traces and the constraint reconstruction~\cite{BatesJones1989}.
For the fixed finite family of stable modes with nonzero horizon values in
section~\ref{sec:budget}, the omitted terms in the future flux cost are
$O[(\sum_j|b_j|^2)^2]$. The corresponding positive quadratic form controls
the mode amplitudes with a constant independent of $\varepsilon$.

\paragraph{Control along the moving horizon.}
The field on a moving horizon changes through both the covariant flux and
radial drift. The tangent vector to the horizon is $v_H=d_++c_H\partial_r$,
where $d_+$ is the outgoing null derivative defined in
eq.~\eqref{eq:nullDefinitions} and $c_H=\dot{\mathcal A}_H/(2\Phi_H\Phi'_H)$. Parallel transport with the
actual gauge potential gives
\begin{equation}
\begin{aligned}
 \widetilde\psi_H&=\mathsf U\psi_H,\qquad
 \mathsf U=\exp\left[-2\ii\int_0^\tau A(v_H)\dd\tau'\right],\\
 \dot{\widetilde\psi}_H&=\mathsf U(\Pi_H+c_H(D_r\psi)_H).
\end{aligned}
\label{eq:horizonTransport}
\end{equation}
The radial term has total variation at most $M_r\Delta\mathcal A$.
It is therefore small even if the drive lasts a long time.

The geometric coefficient in the area law depends on the horizon field and
must be controlled throughout preparation. The focusing constraint gives
$\Phi'\geq1$. Write $\rho=|\psi_H|$
and $B_H=3/2+\rho^2/2-q_H^2/(8\mathcal A_H^2)$.
The horizon equations give
\begin{equation*}
 K_H=\frac{\Phi_H\Phi'_H}{B_H}
 \geq\frac{\sqrt{\mathcal A_i}}{3/2+\rho^2/2}.
\end{equation*}
The variable $F_\rho=\sqrt2\operatorname{arsinh}(\rho/\sqrt3)$ incorporates
this field dependence into the total variation estimate.
Maxwell balance and the Cauchy--Schwarz inequality give
\begin{gather}
 \dot q_H=-4\mathcal A_H\Imag(\widetilde\psi_H^*\dot{\widetilde\psi}_H)
       +8c_H\mathcal A_H\Imag(\psi_H^*(D_r\psi)_H),
 \label{eq:compactMovingChargeBalance}\\
 \operatorname{TV}F_\rho
 \leq\left(\frac{\mathcal T\Delta\mathcal A}
 {\sqrt{\mathcal A_i}}\right)^{1/2}
 +\sqrt{\frac23}M_r\Delta\mathcal A .
 \label{eq:compactMovingChargeTV}
\end{gather}
Here $\operatorname{TV}$ denotes total variation.

Consider a subsequence along which
$\mathcal T\Delta\mathcal A/(\mathcal A_c\varepsilon^2)$ approaches its lower limit.
The bound is immediate if this limit is infinite. If it is finite, the product
is bounded and
$\mathcal T\Delta\mathcal A=O(\varepsilon^2)$.
Eq.~\eqref{eq:compactMovingChargeTV} controls the horizon field and
integrated flux. Together with Maxwell's equation and the area balance, it gives
the estimates below throughout the drive. Substitution into $B_H$ then
recovers the geometric coefficient,
\begin{equation*}
\begin{aligned}
 \sup\rho&=O(\varepsilon),
 &\int_0^{\mathcal T}|\Pi_H|\dd\tau&=O(\varepsilon),\\
 q_H&=Q_c+O(\varepsilon^2),
 &\mathcal A_H&=\mathcal A_c+O(\varepsilon^4),\\
 K_H&\geq\mathcal A_c\kappa_c^{-1}[1-o(1)].
\end{aligned}
\end{equation*}
\paragraph{Charge transfer and the minimum action.}
The accumulated radial drift is controlled by the area increase, while the
entropy cost depends on the squared covariant flux. With normalized time
$s=\tau/\mathcal T$, we define the path directly by
\begin{equation}
 w(s)=1+\frac{1}{\psi_{H,i}}
 \int_0^{s\mathcal T}\mathsf U\Pi_H\dd\tau'.
 \label{eq:fluxPath}
\end{equation}
The difference between $\psi_{H,i}w(\tau/\mathcal T)$ and
$\widetilde\psi_H(\tau)$ has total variation at most
$M_r\Delta\mathcal A=O(\varepsilon^4)$. Substituting $\psi_{H,i}w$ for the transported field in the signed flux
integral therefore changes the integral by
$O(\varepsilon^5)$. The initial charge expansion in eq.~\eqref{eq:chargeRatio}
and the release condition give
$q_{H,\rm off}-q_{H,i}=2\mathcal A_cR|\psi_{H,i}|^2+o(\varepsilon^2)$.
The path has $w(0)=1$ and $w(1)=o(1)$. The integrated Maxwell balance
becomes
\begin{equation*}
 \int_0^1\Imag(\bar w \partial_s w)\dd s=-R/2+o(1).
\end{equation*}
The area law bounds the path action on these subsequences.
An affine endpoint correction makes $w(1)=0$, changing both the action
and the signed integral by only $o(1)$.
Closing the corrected path along the real axis gives a loop with oriented
area $-R/4+o(1)$, equal to half of the signed integral.

To minimize the flux action at fixed oriented area, write
$\mathcal E=\int_0^1|\partial_s w|^2\dd s$ and
$\mathcal I=\int_0^1\Imag(\bar w \partial_s w)\dd s$.
For $0<k<\pi$, introduce a function that vanishes at both endpoints,
$\chi=e^{\ii ks}w-\sin[k(1-s)]/\sin k$.
Completing the square gives
\begin{equation}
 \mathcal E+2k\mathcal I
 =k\cot k+\int_0^1(|\partial_s\chi|^2-k^2|\chi|^2)\dd s
 \geq k\cot k .
\label{eq:compactPathDual}
\end{equation}
Since $\chi$ vanishes at both endpoints, the Dirichlet Poincar\'e inequality
makes the integral nonnegative for $0<k<\pi$.
At exact signed integral $-R/2$, equality requires $\chi=0$.
This gives the circular arc,
\begin{equation}
 w=e^{-\ii ks}\frac{\sin[k(1-s)]}{\sin k},
 \qquad k=\theta/2,
\label{eq:minimumArc}
\end{equation}
where $\theta$ is determined by eq.~\eqref{eq:chargeAction}. The arc has
 action $G(R)$. Applying the same inequality when
$\mathcal I=-R/2+o(1)$ gives $\mathcal E\geq G(R)-o(1)$.

The path definition and endpoint correction give
\begin{equation*}
 \mathcal E=\mathcal T|\psi_{H,i}|^{-2}
 \int_0^{\mathcal T}|\Pi_H|^2\dd\tau+o(1).
\end{equation*}
Combining this relation with the area law, the initial expansions and the
exact entropy balance yields eq.~\eqref{eq:chargeTime}, including preparations
for which the remaining fraction tends to one.

\end{document}